\documentclass[pre,showpacs,showkeys,preprintnumbers,amsmath,amssymb,superscriptaddress,twocolumn]{revtex4}

\usepackage{graphicx}
\usepackage{amsfonts}
\usepackage{amssymb}

\def\rc{\phi_{\rm rcp}}
\def\pj{\phi_J}

\def\pu{\phi_U}

\def\e{\cdot 10^}
\def\e0{\eta_{\rm sol}}

\begin{document}

\title{
Jamming at zero temperature, zero friction, and finite applied shear stress
}

\author{Massimo Pica Ciamarra}
\affiliation{CNISM,
Second University of Naples, 81031 Aversa (CE), Italy}
\affiliation{Dipartimento di Scienze Fisiche,
 Universit\'a di Napoli ``Federico II'' 
 Complesso universitario Monte S. Angelo, 
 Via Cintia, 80126 Naples, Italy}

\author{Antonio Coniglio}

\affiliation{Dipartimento di Scienze Fisiche,
 Universit\'a di Napoli ``Federico II'' 
 Complesso universitario Monte S. Angelo, 
 Via Cintia, 80126 Naples, Italy}
\affiliation{INFN Udr di Napoli and INFM--CNR Coherentia}

\date{Received: \today / Revised version: }

\begin{abstract}
Via molecular dynamics simulations, we unveil the hysteretic nature of the jamming transition of soft repulsive frictionless spheres, as it occurs varying the volume fraction or the shear stress. In a given range of control parameters the system may be found both in a flowing and in an jammed state, depending on the preparation protocol. The hysteresis is due to an underlying energy landscape with many minima, as explained by a simple model, and disappears in the presence of strong viscous forces and in the small $\sigma$ limit. In this limit, structural quantities are continuous at the transition, while the asymptotic values of two time quantities such as the self-intermediate scattering function are discontinuous, giving to the jamming transition a mixed first-order second-order character close to that found at the glass transition of thermal systems.
%We investigate the jamming transition of soft repulsive frictionless particles at zero temperature, in the volume fraction shear stress plane, via molecular dynamics simulations. In a given range of control parameters the system may
%be found both in a flowing and in an jammed state, depending on the preparation protocol. The hysteretic behavior is due to an underlying energy landscape with many minima, as explained by a simple model, and disappears in the presence of strong viscous forces, and in the small $\sigma$ limit. In this limit, structural quantities are continuous at the transition, while two time quantities such as the self-intermediate scattering function, are discontinuous, giving to the jamming transition a mixed first-order second-order character close to that found at the glass transition of thermal systems.
\end{abstract}


\maketitle
Athermal systems such as foams and granular materials undergo a transition from a disordered fluid like state to an amorphous solid, known as jamming transition, as the density is increased and/or the applied shear stress decreased.
The properties of this transition, which is of interest due to its connections to the glass transition of thermal systems, and for its relation to important geophysical phenomena such as avalanches and earthquakes, have not been fully clarified.
Focussing on a system of repulsive soft spheres, Ref.~\cite{Ohern03} showed that the jamming transition at  zero temperature and zero applied shear stress $\sigma$ occurs at a well defined point of the volume fraction $\phi$~\cite{Liu98}, identified with the random close packing volume fraction $\rc$ (point $J$). The jamming transition at point $J$ has a mixed first--order second--order character, whose origin is partially explained by simple geometrical percolation--based models~\cite{Schwarz}. Some quantities, such as the pressure $P$ and the shear modulus $G$, are zero below the transition, and continuously grow as power laws above the transition, while the mean contact number $Z$, which is also zero below the transition, discontinuously jumps to isostatic value $Z = Z_{iso}$ at the transition, and then grows as a power law.  $Z_{iso} = 2D$, where $D$ is the dimensionality, is the theoretical minimum value of the mean number of contacts required for mechanical stability~\cite{Ohern03,Moukarzel,Roux}. Finally, the shear viscosity in the limit of small applied shear stress exhibits a divergence as the jamming point is approached~\cite{Olsson}. 
%How does this scenario change in the presence of a finite value of the shear stress $\sigma$? This question has not yet been fully explored, altough it is expected~\cite{Liu98,Ohern03} that a jamming transition line in the volume fraction $\phi$ shear stress $\sigma$ plane ends continuously at the $J$-point.
It has not yet been fully explored how this scenario changes in the presence of a finite value of the shear stress~\cite{Liu98,Ohern03,Heussinger}. % that a jamming transition line in the volume fraction $\phi$ shear stress $\sigma$ plane ends continuously at the $J$-point.

In this Letter, we investigate via molecular dynamics simulations the jamming transition at a finite value of the applied stress, i.e. in the $\phi$--$\sigma$ plane, discovering that this is actually hysteretic. In a region of the $\phi$--$\sigma$ plane the system may be found both in a flowing and in a jammed state, depending on the preparation protocol. At fixed $\sigma$, the volume fraction at which a jammed system unjams and starts flowing ($\phi_U$) is greater than the volume at which a flowing systems jams ($\phi_J$). A simple one-dimensional model, able to reproduce the observed features, clarifies that the hysteresis is due to the inertia of the system, and to the presence of an underlying energy landscape with many minima.

At small by finite $\sigma$, where the hysteresis is negligible, the jamming transition has a mixed first--order second--order character. Continuous quantities include $P$, $G$ as well as the mean contact number $Z$ (at variance from the $\sigma = 0$ case~\cite{Ohern03}). Discontinuities are found, in close analogy with thermal glass forming systems, in the asymptotic value of the self-intermediate scattering function (the non--ergodicity parameter $f_\infty$) and of others structural relaxation functions. 

{\it Numerical Model} -- We consider a system of spheres of diameter $d$ and mass $m$ in a box of size $l_x = l_y = 16d$ and $l_z = 8d$ (we have investigated values of $l_z$ up to $64d$ to check for absence of finite size effects). We use periodic boundary conditions along $x$ and $y$, while along $z$ the system is confined by rough plates, made by a set of glued particles. The bottom plate is fixed, while the top one (with mass $ml_xl_y/d^2$) is subject to a shear stress $\sigma$.  
Particles interact via the linear spring dashpot model~\cite{Silbert01}. When the center-center separation $r_{ij}$ between particles $i$ and $j$ is smaller than $d$ ($\delta = d-r_{ij} > 0$), particles $i$ and $j$ interact via a repulsive normal force 
\begin{equation}
\label{eq:forcelaw} 
{\bf F}_n = k_n \delta {\bf n} - \gamma_n {\bf v}_{n}
\end{equation}
where ${\bf n} = {\bf r}_{ij}/|{\bf r}_{ij}|$, ${\bf v}_{n} = {\bf n} d({\bf r}_{ij}\cdot{\bf n})/dt$, and $\gamma_n$ chosen fixing the restitution coefficient to $e = 0.88$ (the value of $e$ does not qualitatively change our results). Particles are also subject to a viscous force $-\e0 {\bf v}$ acting at all times, and different values of $\e0$ are explored. Lengths, masses, times, velocities and stresses are expressed in units of: $d$, $m$, $t = \sqrt{m/k_n}$, $v = d_0/t$, $\sigma_0 = k_n/d$. Note that the shear stress is measured in units of the Young modulus of the particles, implying that the hard spheres behavior is recovered in the $\sigma \to 0$ limit.

%\begin{figure}[t!]
%\begin{center}
%\includegraphics*[scale=0.33]{unjam_jam3.eps}
%\end{center}
%\caption{\label{fig:jam}
%(Color online) Velocity of the top plate as a function of the applied shear stress $\sigma$, which is first increased (full lines) and then decreased (dashed lines), at $\e0 = 10$. Different colors refer to different values of the volume fraction, as indicated. At each value of $\phi$, the shear stress at which the system unjams (full vertical lines) when increasing $\sigma$, is greater than the shear stress at which the system jams (dashed vertical lines) when decreasing $\sigma$.
%}
%\end{figure}

The system is prepared via a Lubachevsky--Stillinger~\cite{Lubachevsky,Zhang} like procedure. Particles are placed into the system with infinitesimal radii, rapidly increased to their final value. Then, the system is allowed to relax until the kinetic energy vanishes. As in Ref.~\cite{Zhang}, using periodic boundary conditions in all directions we find $\rc \simeq 0.645$ as the maximum volume fraction at which the system is able to relax in an unjammed (zero pressure) state~\footnote{A corrective term is introduced to take into account the finite size of our sample. In the presence of the rough confining plates, we define the volume fraction as $\phi(N) = N\upsilon/l_xl_yl_z \left[1 + \delta V/(l_xl_yl_z)\right]$, where $\upsilon$ is the volume of a grain, and we have introduced a term $\delta V/l_x l_y l_z$ to take into account the effects of the walls protruding into the system. We have determined $\delta V \simeq 52$ requiring the maximum volume fraction at which the system relaxes in a zero pressure state in the presence of walls to be equal to that obtained without walls.}. 

\begin{figure}[t!]
\begin{center}
{
\includegraphics*[scale=0.33]{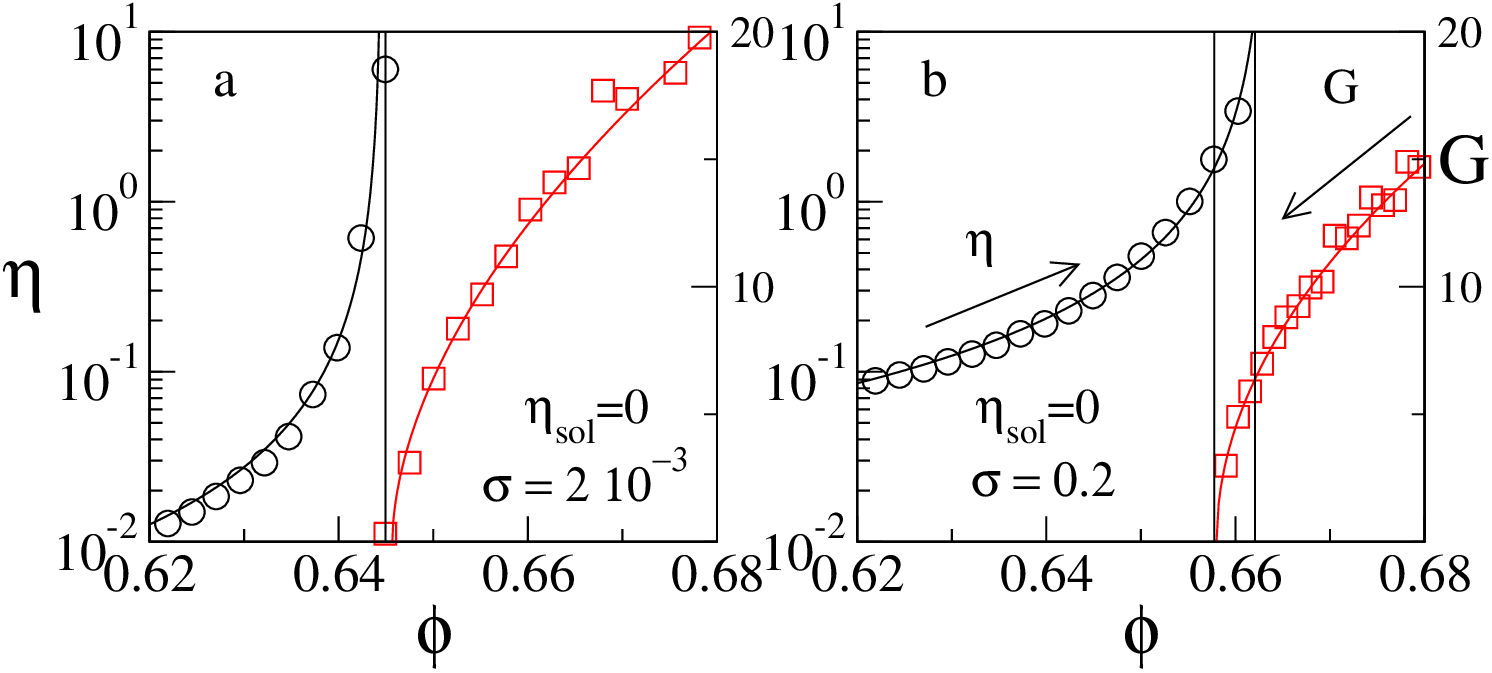}\\
~\hspace{-0.6cm}\includegraphics*[scale=0.33]{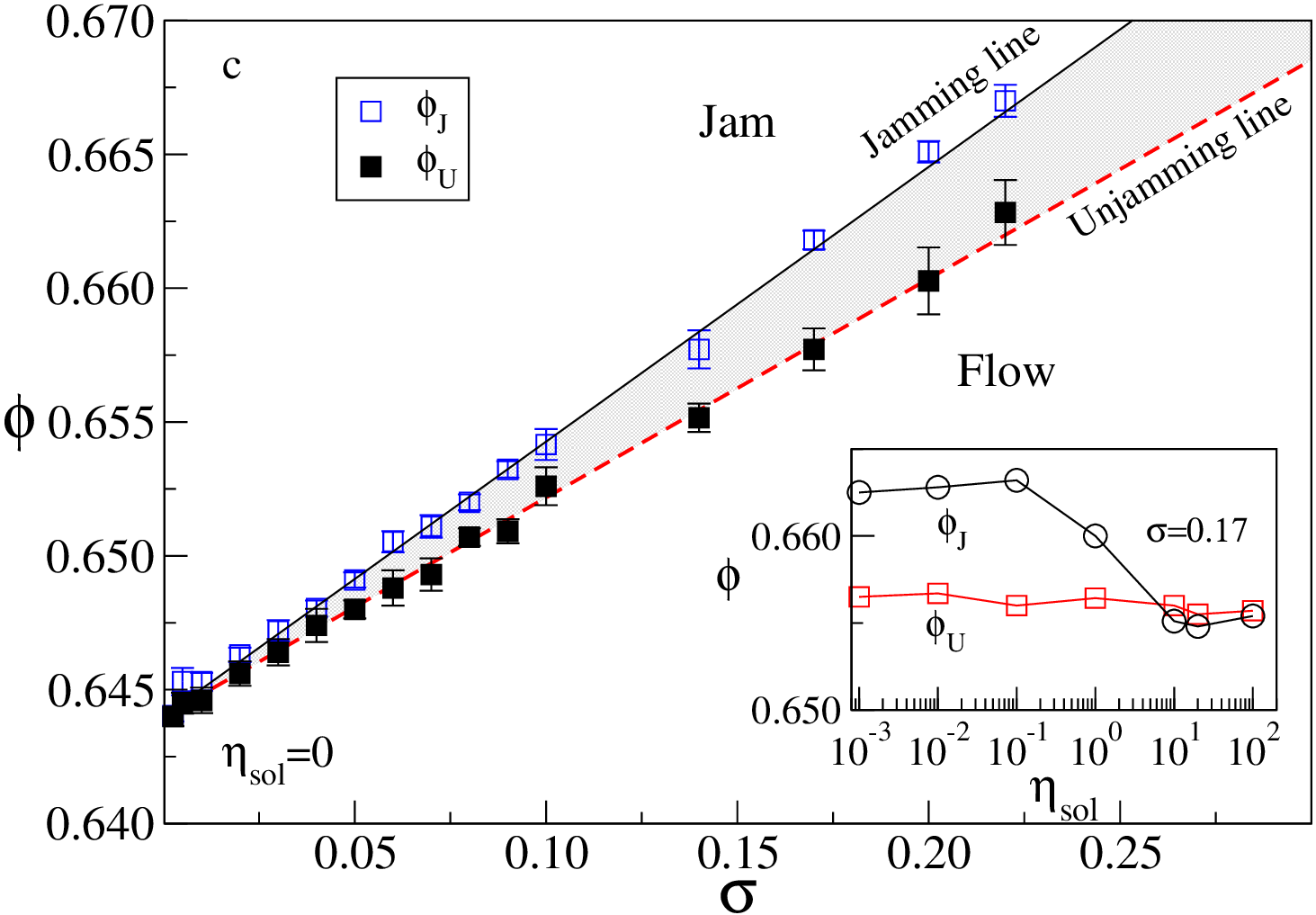}
}
\end{center}
\caption{\label{fig:dia}
(Color online) Dependence of the shear viscosity of the shear modulus on the volume fraction in the presence (panel a, high $\sigma$) and in absece (panel b, small $\sigma)$ of hysteresis. In the prence of hysteresis, the volume fraction at which the viscosity diverges is greater than that at which the shear modulus vanishes. Plain lines are power law fits. %: $\eta \propto (\phi_J-\phi)^{-1.5}$, $G \propto (\phi-\phi_U)^{0.45}$.
Panel (c): jamming phase diagram at $T=0$. Full symbols mark the transition from the jammed to the flowing state (the shear modulus vanishes), while open symbols mark the transition from the flowing to the jammed state (the viscosity diverges). In the gray region, the system has an hysteretic behavior. The hysteresis decreases on increasing the viscosity $\gamma$ of the solvent the particles are immersed in. The inset illustrates that $\pj$ decreases as $\gamma$ increases, until it reaches $\pu$.
}
\end{figure}

{\it Hysteresis at jamming transition} -- 
We have found the response of the system not to be always determined by the volume fraction $\phi$ and by the applied shear stress $\sigma$. For some values of $\phi$ and $\sigma$, it also depends on the preparation protocol. We have therefore considered two limiting protocols, where $\sigma$ is fixed either coming from the jammed or from the flowing phase, respectively. In the first case, the shear stress is slowly increased starting from $\sigma = 0$, until the desired value is obtained. In the other case, $\sigma$ is first fixed to a high value at which the system is seen to flow, and then decreased to its final value.
 %$\sigma$, either fixed coming from the jammed phase (i.e. $\sigma$ is slowly increased starting from $\sigma = 0$), or fixed coming from the flowing phase ($\sigma$ is slowly decreases starting from a high value, where the system is observed to flow).
Once the system reaches the steady state, we have measured the shear viscosity $\eta = v_s/\sigma l_z$, where $v_s$ is the mean velocity of the top plate, and the shear modulus $G$. The shear modulus is $G = l_z \delta \sigma/\delta L$, where
where $\delta L$ is the displacement of top plate of a system of volume fraction $\phi$ jammed under the action of a shear stress $\sigma$, occurring when a perturbing stress $\delta\sigma$ is superimposed. 
%The shear modulus $G$ is:
%\begin{equation}
%G(\phi,\sigma) = \lim_{\delta\sigma \to 0} \frac{l_z\delta\sigma}{\Delta L(\phi,\sigma,\delta\sigma)},
%\end{equation}
%where $\Delta L(\phi,\sigma,\delta \sigma)$ is the displacement of top plate of a system of volume fraction $\phi$ jammed under the action of a shear stress $\sigma$, occurring when a perturbing stress $\delta\sigma$ is superimposed. 

Fig.s~\ref{fig:dia}a,b show the dependence of the shear viscosity and of the shear modulus on the volume fraction, at different values of the shear stress. Each point is the average over $20$ independent runs, and errors are smaller that the symbol size. 
Measures taken coming from the flowing state, see the viscosity to increase with $\phi$, until the system jams at some value $\pj$. At this point, the system is a solid with a finite value of $G$. Measures taken coming from the jammed phase, see $G$ to decrease with $\phi$, until the system becomes a liquid at some value $\pu \le \pj$. At this point, the system starts flowing with a finite viscosity. While at a small value of $\sigma$ (panel a) we found $\pu \simeq \pj$, and higher values of $\sigma$ (panel b) the transition is clearly hysteretic.
%The viscosity increases with the density, until the system jams at some vale $\pj$. %it diverges at a jamming threshold  $\pj$.  At this point, the system is a solid with a finite shear modulus $G$, which increases with $\phi$. When the volume fraction is decreased coming from the jammed phase, the shear modulus decreases until it vanishes at the unjamming threshold $\pu \leq \pj$. At $\pj$, the system start flowing with a finite shear viscosity $\eta$. 
Practically, we have identified $\pj$ and $\pu$ via power law fits of the data shown in Fig.~\ref{fig:dia}a. The vicosity appears to alway diverge as $(\phi_J(\sigma)-\phi)^{-s}$ with $s \simeq 1.5$, while $G$ vanishes as $(\phi-\phi_U(\sigma))^{\gamma}$ with $\gamma \simeq 0.50$, in agreement with previous results at $\sigma =0$~\cite{Ohern03}.% reusltclose to the value found  in the range $0.45$--$0.55$. % ($\gamma$ decreaes with $\sigma$)

We have obtained data as those shown in Fig.~\ref{fig:dia}a,b for different values of the applied shear stress $\sigma$, and we have therefore identified the jamming $\pj(\sigma)$ and unjamming $\pu(\sigma)$ transition lines, which are shown in the $\phi$--$\sigma$ plane of Fig.~\ref{fig:dia}c. We have checked that these lines do not change if estimated at fixed $\phi$ varying $\sigma$. Both lines ends at $\phi \simeq \rc$ (J point) in the $\sigma \to 0$ limit.
The width of the region where hysteresis continuously decreases with the applied shear stress. 
%We have checked that the jamming and the unjamming line do not change if measured not a fixed $\sigma$ varying $\phi$ as we have done, but at fixed $\phi$ fixing the volume fraction, and varying the shear stress (in the quasistatic limit $\dot \sigma = 0$).

The hysteresis depends on the viscosity of the fluid the particles are immersed in $\e0$. In particular, the unjamming transition $\pu(\phi)$ does not depend on $\e0$,  as one may have expected since $\pu$ is determined coming from the jammed phase, where $\e0$ cannot play a role. On the contrary, $\pj$ depends on $\e0$: at a fixed value of the applied shear stress $\sigma$,  $\pj$ decreases as $\e0$ increases, until it reaches $\pu$ and the hysteresis vanishes. The dependence of $\pj$ and $\pu$ on $\e0$ is shown in the inset of Fig.~\ref{fig:dia}c, for $\sigma = 0.17$. 

These results suggest a physical interpretation of the hysteretic behavior. In the region of the control parameters where hysteresis occurs, there are particle configurations (energy minima) which are able to sustain the applied stress. If the system reaches one of these configurations coming from the jammed phase, it has zero kinetic energy, and stays jammed. Conversely, if the system reaches one of these minima coming from the flowing state, it has a kinetic energy large enough to escape from the minima, without jamming.
%with a small kinetic energy, it could be attracted by it (due to the dissipation) and jam. Conversely, if the system visit the configuration with a high kinetic energy, it escapes form it, and keep flowing.

\begin{figure}[t!]
\begin{center}
\includegraphics*[scale=0.33]{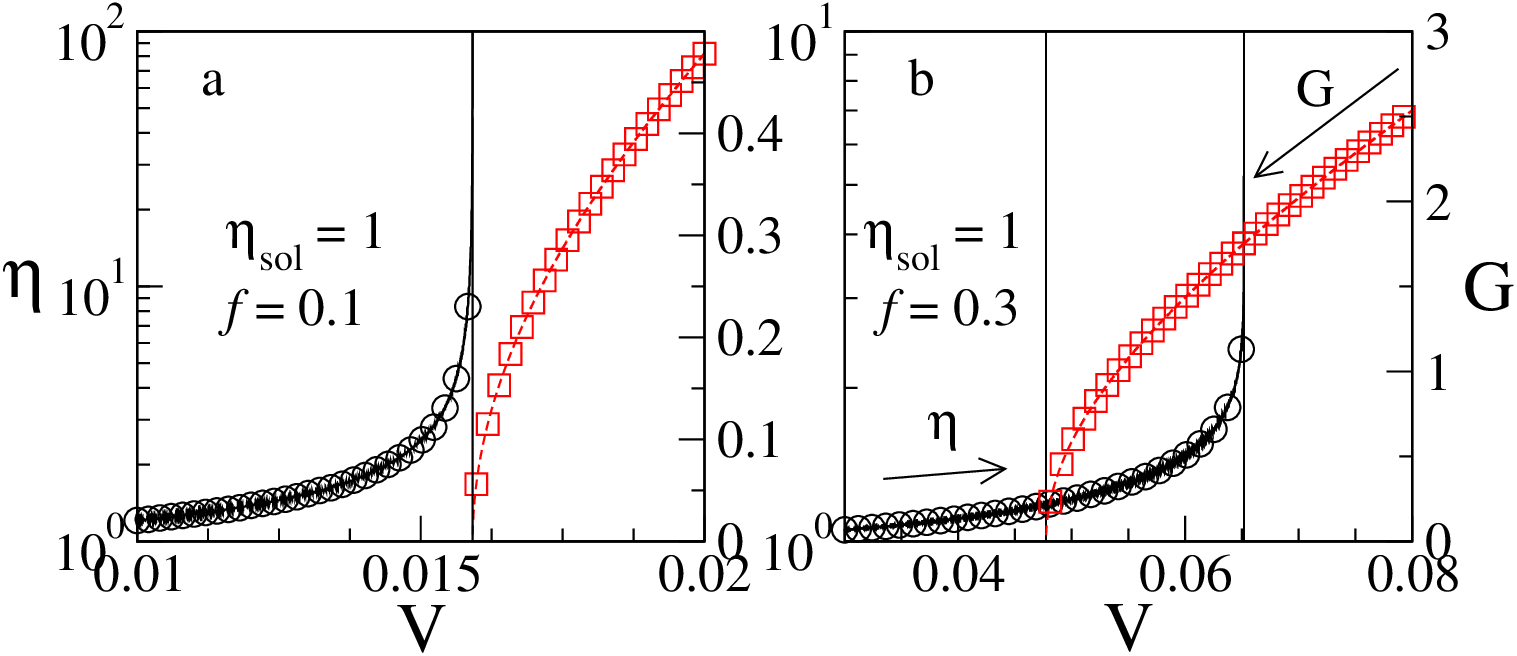}\\
~\hspace{-0.8cm}\includegraphics*[scale=0.33]{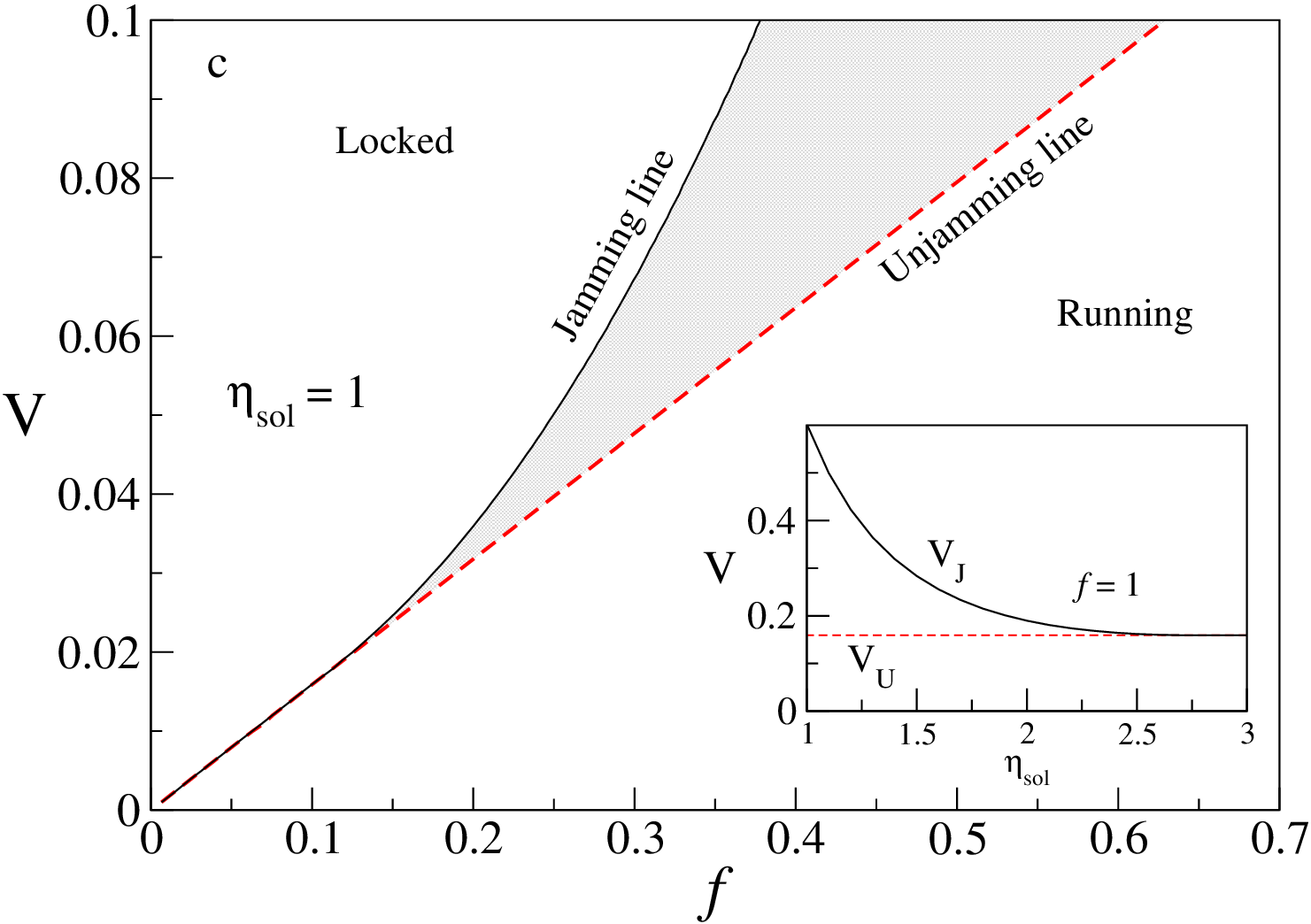}

\end{center}
\caption{\label{fig:dia_model}
(Color online) The same quantities shown in Fig.~\ref{fig:dia}, are here shown as predicted by the model of Eq.~\ref{eq:model}, where $f$ and $V$ play the role of the applied stress $\sigma$ and of $\phi-\rc$, respectively. The model reproduces the hysteretic behavior, as well as the dependence of the width of the hysteretic region on both the applied stress and the viscosity of the solvent $\e0$.
}
\end{figure}

{\it Model} -- The hysteretic behavior observed at the jamming transition, and its dependence on $\e0$, are qualitatively explained by a simple model for the motion of a particle in a energy landscape with many minima $W(x)$, subject to a constant driving force $f$ and to a viscous one $-\e0 \dot x$:
\begin{equation}
\label{eq:model}
 m \ddot x = f - dW(x)/dx - \e0 \dot x.
\end{equation}
For simplicity sake, here we use $W(x) = -V\cos(\omega x)$, and fix $m = 1$ and $\omega = 2\pi$. $V$ measures the depth of the energy minima, and we therefore relate it to the volume fraction ($V$ increases as $\phi$ overcomes $\rc$).
With this choice for $W(x)$, Eq.~\ref{eq:model} has been extensively investigated as it models a large variety of physical processes (see Ref.~\cite{Risken} for a comprehensive review). 
Eq.~\ref{eq:model} admits two type of solutions, known as `running' (the mean velocity is $\langle \dot x \rangle > 0$) and as `locked'($\dot x = 0$), which correspond to our flowing and jammed state, respectively. In the running state, it is possible to define the shear viscosity as $\eta = \langle \dot x \rangle/f$. In the locked state, the equilibrium position $x_{eq}$ is fixed by $f = \left. dW(x)/dx\right|_{x_{eq}}$, and the shear modulus is $G(f,V) = \left. d^2W/dx^2 \right|_{x_{eq}}$.
Eq.~\ref{eq:model} reproduces the hysteretic behavior observed across the jamming transition because in a given of the control parameters $V$ and $f$ both the running and the locked solution are allowed, the observed one depending on the preparation protocol. For instance, we show in Fig.s~\ref{fig:dia_model}a,b the dependence of the shear viscosity and of the shear modulus on $V$. The shear viscosity, measured when $V$ increases, diverges at a jamming threshold $V_J$, while the shear modulus, measured decreasing $V$, vanishes at an unjamming threshold $V_U \le V_J$. This behavior closely resemble that of Fig.~\ref{fig:dia}(a). 

From Eq.~\ref{eq:model} it is also possible to determine the `jamming diagram' in the $V$--$f$ plane. The unjamming line is that where the shear modulus vanishes, $G(f,v_0) = 0$ (i.e. $V_U = \omega f $), while the jamming line (numerically determined) is that where the shear viscosity diverges. The resulting diagram, which is shown in Fig.~\ref{fig:dia_model}(c), qualitatively reproduces that of Fig.~\ref{fig:dia}(c). In particular, the hysteretic region decreases as $f$ decreases. Moreover, as shown in the inset, Eq.~\ref{eq:model} also reproduces the dependence of the width of the hysteretic region on the shear viscosity.  

%There are some qualitative differences between the results of the numerical simulations shown in Fig.~\ref{fig:dia}, and the predictions of the model shown in Fig.~\ref{fig:dia_model}. One of the sources of these discrepancies is the different role played by $\e0$.  While in the model the damping term proportional to $\e0$ is the only dissipation mechanism, in the numerical simulations dissipation is always present due to the inelastic interaction between the grains. The limit $\e0 \to 0$ is therefore expected to be different in the two cases, as observed comparing the insets of Fig.~\ref{fig:dia}(b) and Fig.~\ref{fig:dia_model}(b).

\begin{figure}[t!]
\begin{center}
\includegraphics*[scale=0.33]{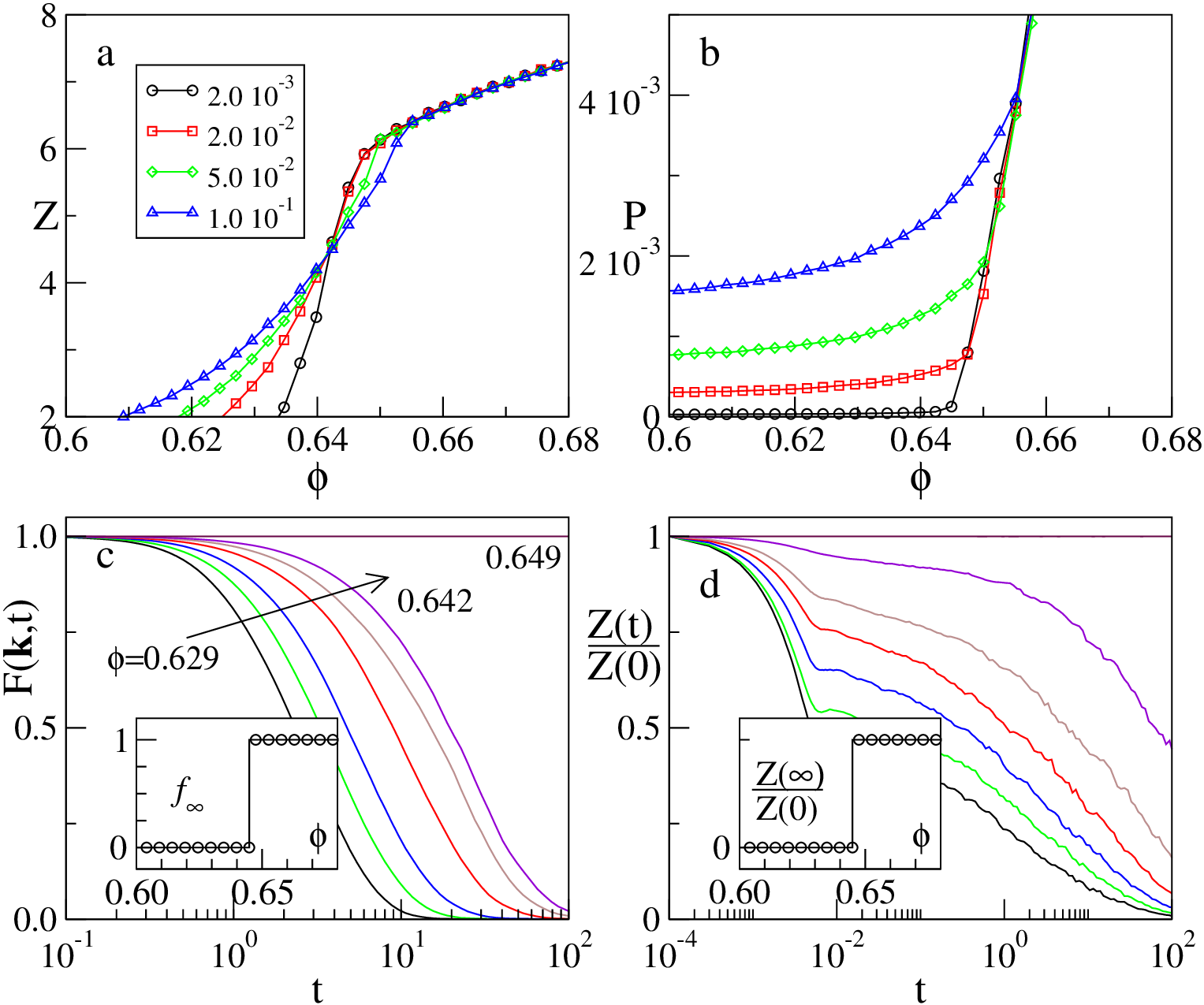}
\end{center}
\caption{\label{fig:zpg} (Color online) 
Structural changes at the jamming transition at small $\sigma$. Panel (a) and (b): dependence of the mean contact number and of the normal pressre on the volume fraction, for different values of $\sigma$, as indicated. Panel (c): self-intermediate scattering function for different values of $\phi$ (main panel), and volume fraction dependence of its asymptotic value $f_\infty$ (inset). Panel (d): two-time mean contact number $Z(t)/Z(0)$ (main panel) for different values of $\phi$ (same as in panel c), and $\phi$ dependence of its asymptotic value $Z(\infty)/Z(0)$ (inset). In panels a,c and d, $\sigma = 2~10^{-3}$.}
\end{figure}

{\it Structural signatures of the jamming transition} --
We now consider how dynamical and structural quantities of the system changes at the jamming transition.
We focus on small values of the applied shear stress $\sigma < 0.1$ where, as shown in Fig.~\ref{fig:dia}(b), hysteretic effects are negligible. %at $\e0 = 0$. At shown in Fig.~\ref{fig:zpg}(a) shows that for $\sigma$ small enough the shear viscosity diverges at the same volume fraction where the shear modulus vanishes.
We show in Fig.~\ref{fig:zpg}(a) the dependence of the mean contact number on the volume fraction: $Z$ continuously increases with $\phi$, and has a cusp at the transition. Above the transition, the effect of the applied shear stress on $Z$ is negligible, and $Z$ grows as a power law with exponent $\simeq 0.5$ as at $\sigma = 0$~\cite{Ohern03}. The pressure $P$, shown in Fig.~\ref{fig:zpg}(b) also increases with $\phi$, linearly above the jamming threshold.
Note that, due to the continuous collisions of the flowing grains, below the jamming transition both the pressure and the mean contact number have finite values, at variance with the $\sigma = 0$ case~\cite{Ohern03}. The $\sigma = 0$ case is approached continuously as $\sigma$ decreases. For instance, we have evaluated via a polynomial fitting the left derivative of $Z$ at $\pj$, $\lim_{\phi \to \pj} \partial Z/\partial\phi$, finding that it diverges as $\sigma^{-x}$, $x \simeq 0.2$, as $\sigma$ decreases.

While $Z$, $G$ and $P$ (not shown) are continuous at the transition, there are quantities which are discontinuous. Precisely,
the asymptotic values of some two-time correlation functions are discontinuous, as in glass forming systems. Two of them are shown in Fig.~\ref{fig:zpg}b and Fig.~\ref{fig:zpg}c. The first one is the self--intermediate scattering function $F({\bf k},t) = \left< \Phi({\bf k},t)\right>$, where $\Phi({\bf k},t) = 1/N\sum_j \exp(-i {\bf k} \cdot ({\bf r}_j(t)-{\bf r}_j(0)))$, we have investigated in the direction ${\bf k} = (0,2\pi/d,0)$ perpendicular to both the confining plates and to the driving force. 
% (in order to avoid effects due to the presence of both the shear stress, as well as of the confining plates). 
$F({\bf k},t)$ goes to zero in the flowing regime, while it stays to one when the system is jammed. The asymptotic value of $F({\bf k},t)$, known as non--ergodicity parameter $f_\infty$, is therefore discontinuous at the transition. 
%With respect to the behavior observed in many glass forming, in athermal systems the so-called $\alpha$ relaxation, which is due to cage motion, is missing~\cite{Hatano}. 
The second quantity is the two--time mean contact number 
\begin{equation}
\label{eq-zt}
Z(t) = \frac{1}{N}\sum_{ij} \langle c_{ij}(t) c_{ij}(0)\rangle,
\end{equation} 
where the sum runs over all particles, and $c_{ij}(t) = 1(0)$ if particles $i$ and $j$ touch (do not touch) at time $t$. $Z(0)$ is the usual mean contact number. As shown in Fig.~\ref{fig:zpg}d, $Z(t)/Z(0)$ goes to zero when the system flows, while stays to one when the system is jammed. Its asymptotic value is therefore discontinuous at the transition, exactly as $f_\infty$. The analogy with the glass transition is also validated by the study of the relaxation time $\tau$ ($F({\bf k},\tau) = 1/e$), which we have found to diverge with the shear viscosity $\tau \sim \eta \sim (\phi_J-\phi)^{-s}$ ($s \simeq 1.5$) as the transition is approached, and by the study of the dynamical susceptibility $\chi_4$, defined as the fluctuation of the self-intermediate scattering function, having a maximum $\chi^*$ at a time $t^*$, with $t^* \propto \eta$, and $\chi^* \propto (\phi_J-\phi)^{-\nu}$, $\nu \simeq 0.8$.
Since our system relaxes at densities which are close to $\rc$, shear appear to be much more efficient than thermal motion in inducing the structural relaxation of the system. The scenario of Fig.~\ref{fig:zpg}, which we have found at small $\sigma$ and $\e0 = 0$, is also observed at high $\sigma$ provided that $\e0$ is large enough for the hysteresis to be negligible.

{\it Discussion} -- 
The hysteretic nature of the jamming transition in the $\phi$--$\sigma$ plane implies the presence of memory, not predicted in the frictionless case considered here. 
%has no counterpart in the jamming transition of thermal system in the $\phi$--$T$ plane, suggesting that the analogy between temperature and shear stress brakes down at large $\sigma$. In particular, the hysteresis implies the presence of memory, unexpected in the frictionless case considered here. 
At small but finite $\sigma$, where the hysteresis is negligible, the jamming transition has a mixed first--order second--order character which is close to that observed in glass forming systems. One time quantities, including the mean contact number, are continuous at the transition, but the asymptotic value of two-time correlation functions, such as that of the self-intermediate scattering function (the non--ergodicity parameter) or that of the two--time mean contact number, is discontinuous.
The jamming transition at $\sigma = 0$ appears therefore the $\sigma \to 0$ limit of a glassy-like transition observed at all values of $\sigma$.
Important open questions ahead include the understanding of the role of frictional forces, which must be taken into account to describe real granular systems~\cite{Zhang,Dauchot,Somfai,Grebenkov,noi}, as well as that of a finite temperature~\cite{Nagel,Witten}.

We thank A. Fierro and M. Nicodemi for interesting discussions, and acknowledge computer resources from CINECA, Unina--SCOPE Grid, INFN--Grid and CASPUR.

\end{document}